# Artificial Nanophotonic Neuron with Internal Memory for Biologically Inspired and Reservoir Network Computing

David Winge[1,3], Magnus Borgström[1,3], Erik Lind[2,3] and Anders Mikkelsen[1,3,*]

[1]*Department of Physics, Lund University, Box 118, 22100 Lund, Sweden*
[2]*Department of Electrical Engineering, Lund University, Box 118, 22100 Lund, Sweden*
[3]*Nano Lund, Lund University, Box 118, 22100 Lund, Sweden*

## Abstract

Neurons with internal memory have been proposed for biological and bio-inspired neural networks, adding interesting functionality. We propose and model a nanoscale optoelectronic neural node with charge-based time-limited memory and signal evaluation. Connectivity is achieved by weighted light signals emitted and received by the nodes. The device is based on well-developed III-V nanowire technology, which has shown high photo-conversion efficiency, low energy consumption and sub-wavelength light concentration. We create a flexible computational model of the complete artificial neural node device using experimental values for wire performance. The model can simulate combinations of nodes with different hardware derived properties and widely variable light interconnects. Using this model, we simulate the hardware implementation for two types of neural networks. First, we show that intentional variations in the memory decay time of the nodes can significantly improve the performance of a reservoir network. Second, we simulate the nanowire node implementing an anatomically constrained functioning model of the central complex network of the insect brain and find that it functions well even including variations in the node performance as would be found in realistic device fabrication. Our work demonstrates the feasibility of a concrete, variable, nanophotonic neural node with a memory. The use of variable memory time constants to open new opportunities for network performance is a general hardware derived feature and should be applicable for a broad range of implementations.

Subject areas: Optoelectronics, electronics, nanophysics

[*] Corresponding author. anders.mikkelsen@sljus.lu.se

# I. INTRODUCTION

In natural and artificial neural networks, memory and computation are combined. This is usually accomplished by building memory into the weights of the neural connections, while the individual neurons evaluate the incoming signals from all its connections to produce an output. However, in both biological as well as artificial neural systems several examples can be found of the neurons themselves possessing an internal memory [1–3]. Giving the neurons a memory significantly increases the context that the network can remember and thus the assignments it can accomplish [4,5]. In biologic systems, several examples have been found in which the neurons appear to have a memory [3,6,7]. Another recent example is for networks derived from anatomically constrained models of the central complex of the insect brain [8]. Here neurons with a leaky memory are introduced as an important part in enabling the network to track the position of the insect's nest while foraging and then subsequently returning to the nest. While the memory effect of individual neurons of this model could be carried out by a collection of intercommunicating neurons, it was clear from the modelling that one type of neuron with an internal memory could perform the necessary navigation task. Thus, it is of interest to propose and investigate artificial neuromorphic neurons with a memory. One prominent example found in artificial networks using so called long short-term memory [4,9]. Here the active memory neural node component is constructed using several neurons that are communicating in a loop which creates a leaky memory (time constant for leaking can then be set by connection weights).

For bio-inspired neural networks the main energy expenditure and complexity is in the communication between components. Traditional transistors operate very efficiently, but the additional energy cost in communication is high and increasing connectivity is difficult in their planar geometry [10–13]. The use of light can be highly advantageous, as optical photons carry little energy, communication with light is fast and high information density can be encoded using wavelength, polarization, and intensity [10,14]. Photonic solutions based on existing technologies have indeed shown great progress in recent years for neuromorphic computing [10,15–19]. However, their physical footprint is very large (limited by the wavelength of light) and while energy expenditure can in principle be extremely low it is often limited by losses by inevitable optical-electronic conversion. That this can be solved using nanostructures has been acknowledged, but not realized [20]. III-V semiconductor nanowires (NWs) represents one of the most mature nanotechnology platforms, with a very broad range of available structures with unique and record-breaking performance for electronics and photonics [21–27]. Recently it has been shown that III-V NWs can be used as hardware for an artificial neural network [28] in which the weighted connectivity between nodes is achieved by overlapping light signals inside a shared quasi 2D waveguide – a broadcasting concept. This decreases the circuit footprint by two orders of magnitude compared to existing optical solutions. The evaluation of optical signals is performed by neuron-like nodes constructed from efficient III–V nanowire optoelectronics. This minimizes power consumption of the network [28].

In this work, we propose and explore the function of a III-V nanostructured neural network node with a charge-based memory based on NW photodiodes and LEDs. The optical interconnect concept of this hardware implementation was successfully simulated previously for a bioinspired network [28]. A detailed quantitative design of the node is proposed and described based on realistic geometric parameters for the NWs. The node device consists of three interconnected NWs that are simulated using benchmarked experimental values. The nodes communicate (instantaneously) using light in the near infrared range. These results are used to build a general neural network model that can simulate wide ranging node connectivity and performance. We use the inherent opportunity for variation in node memory lifetime to implement a reservoir network with substantially better performance than a standard network with identical nodes. We find that we can realize the navigation complex of the insect brain using realistic nanowire parameters including errors and noise in the network. While the present work demonstrates a particular physical implantation of a nanophotonic network, the concepts are likely applicable to a range of physical implementations which offer variability in memory properties and network connectivity.

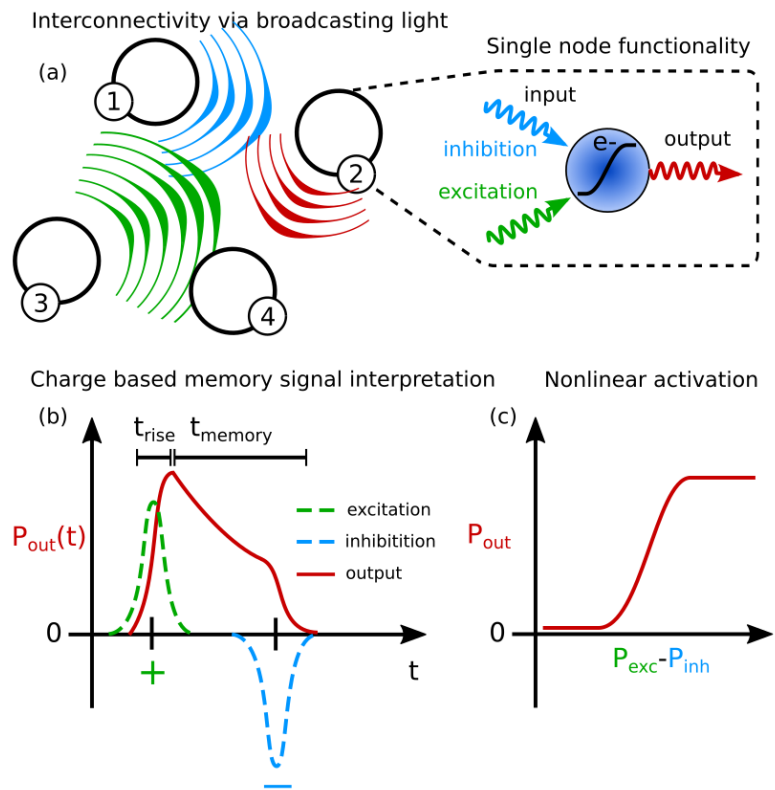

FIG 1: Basic functionality required of the neural node nanodevices for network computations. (a) Schematic illustration of a set of connected nodes communicating with a broadcasting scheme using optical signals that inhibit or excite the nodes. The internode coupling strengths are given by local light intensity and absorption efficiency at the receiver units. The signal interpretation in the node is done electrically after conversion of the light into an electrical current. (b) Schematic example of the desired (electrical) impulse response of a dynamic node upon (light) excitation and inhibition. An exciting pulse pushes the node into an active state. This state will gradually decay with memory timescale $\tau_{MEM}$ leading to a reduced output. An inhibiting pulse that arrives later will subtract from the remaining value of the active state which will lead to a faster termination of output activity (if the inhibiting pulse is strong enough). (c) Typical nonlinear activation function needed for signal processing in each node.

## II. NEURAL NODE DESIGN

### A. Overall concept and design required of the neural node

We first describe the concept and node design in general terms, the functionality needed in such a device and the properties that this requires. The aim is a nano-optoelectronic device that communicates with light, while signal processing and memory function is electrical as illustrated in Fig 1(a). In the present work, the focus is on exploring the functionality of a NW neural node, especially the inclusion of a charge-based memory. Previous work described the optical coupling between III-V NW based neural nodes with geometrically weighted optical interconnects [28]. The basic functionality of the neural node with a memory can be summarized by the capability of receiving, comparing, and transmitting light signals. In the simplest case, a node must compare two different signals, where one is interpreted as excitatory and the other as inhibitory following a sigmoid function as in Fig. 1(c). If these signals arrive at the same time, and the inhibitory signal is similar to (or larger than) the exciting, no net signal is produced. If the exciting signal arrives first and there is a sufficient time delay between the signals, an output signal will be generated for a time (depending on the memory decay $\tau_{mem}$) and will only be terminated when the inhibiting signal arrives, as exemplified in Fig. 1(b). This ability to compare excitation and inhibition is central to allow a neural network consisting of such nodes to perform computations. The memory of the network is then defined by how long the system remembers the exciting and inhibiting signals, thus the memory decay time $\tau_{MEM}$. In general, by tuning the memory decay timescale, the node can either perform as a signal amplifier with a short memory timescale, passing on signals as quickly as its bandwidth allows, or as an integrator, storing information over time using a long memory timescale. The speed with which to transmit signals is fundamentally limited by the rise time of each node. In our optoelectronic device design presented below, these timescales are all governed by electrical processes in the node (fraction of ns), while the optical transmission time is much faster in comparison (order of fs) thus the timescale of the communication itself can be seen as instantaneous.

### B. Node realization using III-V semiconductor NWs

We now show how the necessary functionality can be achieved by connecting multiple NWs with specific functionality in a planar geometry as depicted in Fig. 2(a). The device function can be described in a circuit diagram as shown in Fig. 2(b): Two NW photodiodes are stacked in the left part of the device, sensitive to two different wavelengths (exciting or inhibiting) that generate currents of opposite polarity with respect to memory in between. These photodiodes can be grown as separate NWs or together with a low-doped separation layer, to avoid tunnelling between the two pin-diodes. At the memory capacitor, these excitation and inhibition currents are effectively compared and the charge resulting from the net current is stored at $C_{MEM}$. The resulting floating gate potential $V_{gate}$ links the receiver part with the transmitter NW electronically through the centre NW consisting of a wrap-gate Field Effect Transistor (FET). If $V_{gate}$ is larger than the transistor threshold voltage $V_T$, the LED in the transmitter NW (active region in the transmitter NW), turns on and sends out a signal in the network.

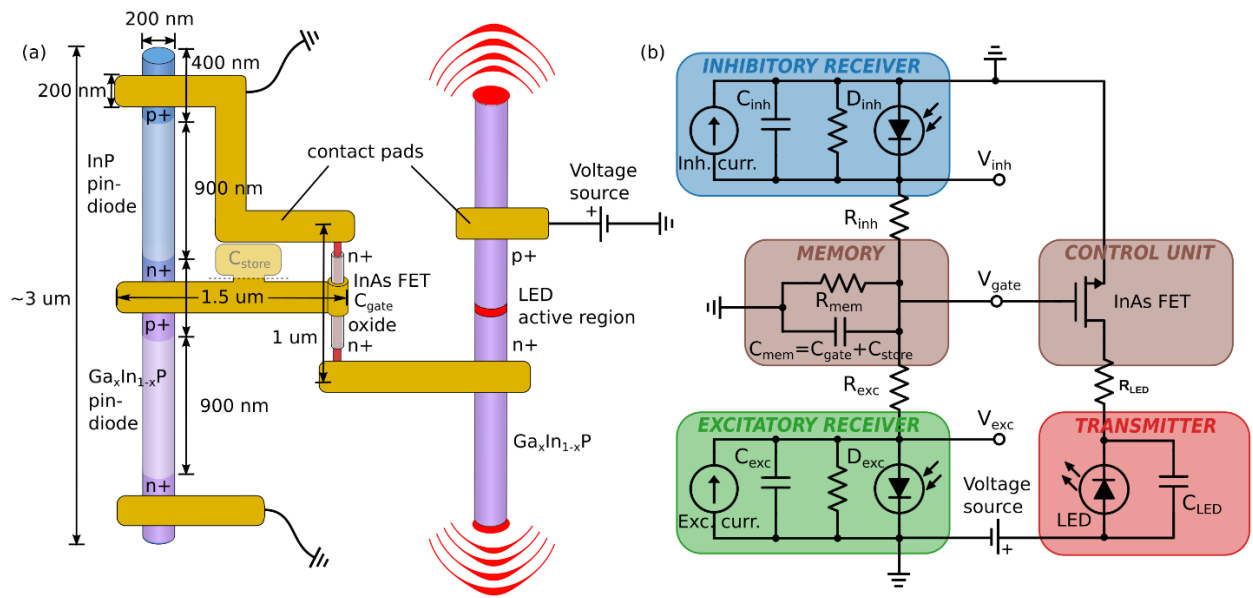

FIG. 2: Optoelectronic node realized by NW design. a) Realization using specific NW geometry with specific compositions and sizes of the components and connections in the circuit. b) Equivalent circuit diagram for the NW device in a). Left leg holds the receiver parts and the floating gate memory, whereas the right leg holds the control unit and the transmitter. $R_{mem}$ does not have a direct correspondence in (a) but encapsulates all pathways of charge leakage from the memory.

Fig. 2(a) display the selected semiconductor materials and geometry for the presented design. These are not the only possible options but chosen here for their already proven characteristics. $Ga_xIn_{1-x}P$ is a well-known direct-bandgap material (for fractions of Ga up to x=75% [29]), useful both for optical absorption and emission. The same pin-diodes as in our design are achieved repeatedly in InP NW solar cells. GaInP has been used in NW LEDs in the past [30,31] in order to realize core-shells structures. Here we assume an axial junction for simplicity, although this is not crucial to our proposal. For the transistor, InAs is chosen as NW FETs in this material system are routinely fabricated and show excellent characteristics [32].

From the equivalent circuit diagram in Fig. 2(b), several relevant timescales can be identified from the circuit subparts. Starting from the receiver part on the left-hand side, the response time of the gate voltage $V_{gate}$ (following an absorbed optical signal) will depend on the series resistances $R_{inh}, R_{exc}$ connecting the memory and photodiodes as well as the capacitances $C_{inh}, C_{exc}$ of the respective photodiode. The memory timescale on the other hand, will be set by the dominant leakage channel from $C_{mem}$, which is represented in the model by $R_{mem}$, resulting in the memory time scale $\tau_{mem} = R_{mem}C_{mem}$.

The dynamic parameters of the photodiodes were extracted from finite element modelling of the light induced current in the NWs. The transport model was calibrated against experimental data from solar cell NW arrays as shown in Fig 3(a). In order to reach a good agreement, the unintentional doping in the intrinsic segment was varied in the simulations. Results from the calibrated NW photodiode model are shown in Fig. 3(a) for different values of the surface recombination velocity, which is the main non-radiative decay channel for minority carriers in our simulations [33]. Using our calibrated NW photodiode model, the response to a light pulse was modelled as shown in Fig. 3(b) and from this the fall times of 20 ps, 20 ps, and 10 ps were extracted for the surface recombination velocities $S$ of $10^4$, $10^5$ and $10^6$ cm/s, respectively. The curve for $10^4$ cm/s is biexponential to a higher degree as compared to the other values of $S$, with the slow part decaying with a timescale of 200 ps, explaining the difference in appearance. This is a consequence of the competing physical processes of charge extraction and recombination. The value of $S=10^5$ cm/s was used for the pin-diode our device simulations below as it provides a good fit to the experimental data. Using the relation $\tau = RC$, the capacitance was calculated from an estimate of the series resistance of the pin-diode (see SI).

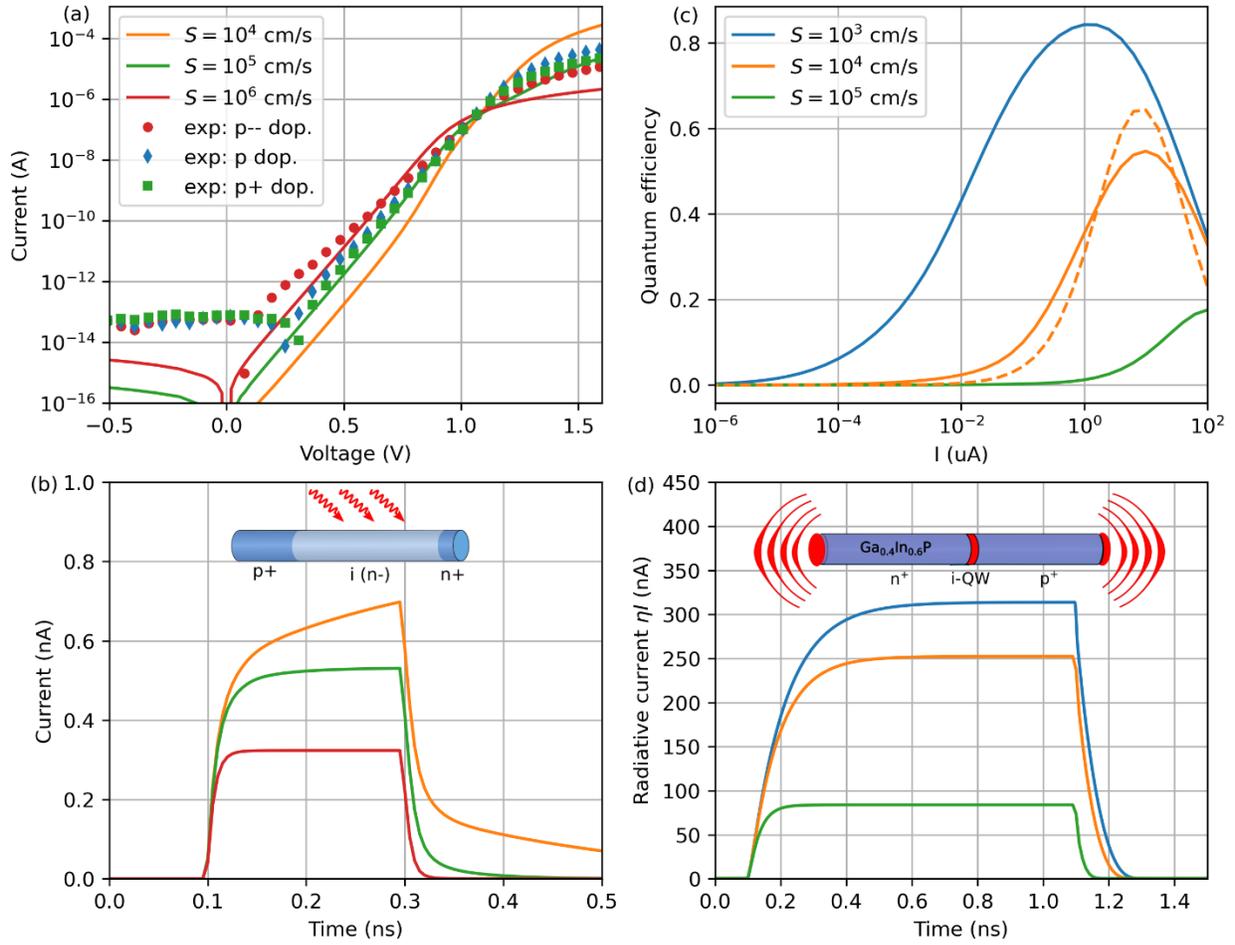

FIG. 3: Modelling of optoelectronic NW devices based on experimental data. (a) Dark current versus voltage for the pin-diode. Results for the NW model (lines) with different surface recombination velocities, are compared to experimental results (markers) from [33] where the doping in the p+-segment was varied as indicated in the legend. (b) Calculated time-resolved photo-current for the pin-diode. (c) Internal quantum efficiency of the NW LED. Dashed line indicates a two-parameter fit to a simple ABC-model, which was then used as input to the dynamic neuron model. (d) Calculated time-resolved photo-emission for the NW LED.

In the right leg of Fig. 2(b), there are two additional timescales: the delay time of the FET responding to changes in $V_{\mathrm{GATE}}$ and the response time of the LED to changes in the FET source-drain current. The delay time of the FET is drastically shorter than the LED response time, with typical values of picoseconds versus a fraction of nanoseconds, respectively. This allows us to model the current through the LED as a simple RC circuit driven by the resulting source-drain current of the NW FET. In order to extract the response time of the LED, we use the same finite-element transport model as in Fig. 3(a), and model a quantum well LED incorporated into a NW. The internal quantum efficiency and the response to a current pulse are shown in Fig. 3(a) and 3(b), respectively. Again, the surface recombination velocity was used as a parameter and as seen in Fig 3(a) it strongly affects the radiative recombination efficiency in the NW LED. From the pulse response we extract rise times of 110 ps, 90 ps and 30 ps for the surface recombination velocities of $10^3$, $10^4$ and $10^5$ cm/s, respectively. The fall times were shorter in general due to the carrier emptying effect as charges are pulled out of the quantum well by the changing electric field [34]. For our device simulations, we use the data for $S=10^4$ cm/s for the LED.

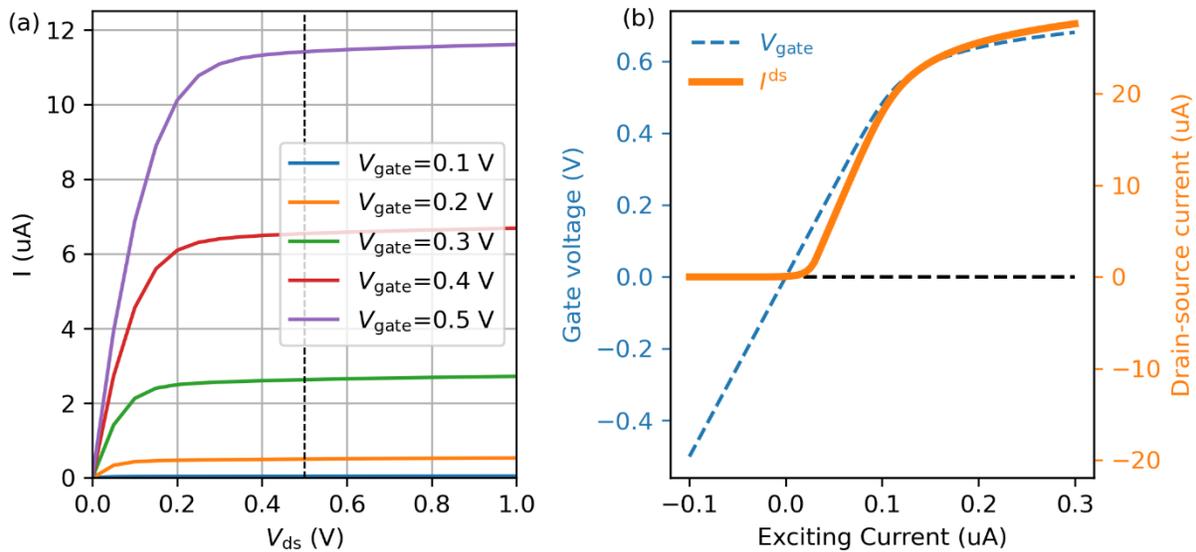

FIG. 4: Modelled transistor and resulting activation function calibrated against experimental data. (a) Simulated source-drain voltage sweep for a set of different gate voltages for the designed NW field effect transistor. Minimum $V_{ds}$ during the applications is marked by a dashed line, which is also the voltage used to extract the source drain current $I^{ds}$ as a function of $V_{gate}$. (b) Stationary activation function in terms of the source-drain current of the transistor, derived from the excitatory part of Fig. 2(b) using typical device parameters (see Appendix A). The dashed blue line indicates the resulting gate voltage as a function of exciting current. This drops off as the photodiode is eventually pushed above threshold.

Due to its fast dynamics compared to the NW, the transistor can be modelled using a time-invariant approach. In order to extract the relevant parameters, an additional finite element model was developed to model a wrap-gate InAs NW FET with a diameter of 50 nm and a gate length of 200 nm (see SI). This model was calibrated against experimental data of samples similar to those in [32], as shown in Fig. 4(a).

Using the transistor model in combination with the physical models for the receiver photo-diodes, a stationary activation function can be calculated (see Appendix A), mapping *input* currents in the photo-diodes to source-drain *output* currents of the transistor, as shown in Fig. 4(b). This shape closely resembles a sigmoid activation function but more importantly carry two distinct nonlinear features: at low input current it acts as a rectifier while it provides diminishing increases of current above a certain threshold. In between these extremes the output current varies linearly with the input. It is an interesting feature that the calibrated transistor model allows for tuning of the activation function in terms of bias and slope, by varying, for example, gate length and using different gate metals. This is a relevant feature for the realization of general neural architectures.

In addition to providing the design with the crucial non-linear element, the transistor acts as a current source to the transmitter line in the right part of Fig. 2(a,b). This allows the system to generate the required gain to propagate the signal through the physical network, as coupling strengths will be far below unity which is standard in model neural architectures.

The node device in Fig. 2 is naturally operated in a free-running mode, meaning that the driving bias is constant and the signals themselves supply the dynamic input. The absence of a clock frequency or duty cycle operation simplifies the implementation. Another good property is that the device can be driven with a very limited number of bias connections. A connection to ground is necessary, which can be shared by all devices in a network, and a driving voltage is needed to generate the current through the transmitter NW. It is thus possible to power a network through a single biasing contact per node device (in addition to the common ground) which limits the number of contacts needed in our design.

The speed with which the node can collect and transmit signals again is controlled mainly by the two longest timescales, the memory RC constant and the LED rise time. Stray capacitances are present in the system (but not shown explicitly in Fig. 2(a,b)) for example between the charge storage site and the contact pads, increasing the memory capacitance and further limiting the speed (see SI). When designing a memory node, the storage capacitance $C_{MEM}$ can be deliberately increased, as shown with a shaded gold pad in Fig. 2(a). This, alongside tuning of the memory resistance $R_{MEM}$, provides the freedom to design the timescale on which the activation of the memory node is sustained.

### C. Time dependent model of the node device and network

In this section we provide a dynamical model for each node and for the complete network. This is derived under the following assumptions: the gate voltage under operation is far from the threshold voltages of $D_{inh}, D_{exc}$, the delay time of the transistor is negligible in comparison to the LED response time, and the transit time of the optical signals between nodes inside the physical network (order of fs) is negligible compared to the electronic timescales (order of ns)

From the diagram in Fig. 2 we can identify a set of time-dependent variables. The receiver is described by the dynamic voltages $V = (V_{\text{inh}}, V_{\text{exc}}, V_{\text{gate}})$ that are updated according to

$$\frac{dV}{dt} = AV(t) + BI(t) \tag{1}$$

where $A$ and $B$ are matrices describing the internal system dynamics (see Appendix B), in terms of the RC constants and capacitances indicated in Fig. 2(b). Especially important is $\tau_{\text{mem}} = R_{\text{mem}} C_{\text{mem}}$ which govern the memory loss rate in the near-steady state when $V_{\text{inh}} = V_{\text{exc}} = V_{\text{gate}}$. Furthermore, $I(t) = (I^{\text{inh}}(t), I^{\text{exc}}(t), 0)$ are the input currents that are generated in the photodiodes upon excitement, either from external sources or connected nodes.

The current through the transmitter LED, denoted $I^{\text{LED}}$, is modelled using a standard RC circuit model with

$$\frac{dI^{\text{LED}}}{dt} = \frac{1}{\tau_{\text{LED}}} \left( I^{\text{sd}}\left(V_{\text{gate}}(t)\right) - I^{\text{LED}}(t) \right) \tag{2}$$

Where $I^{\text{sd}}$ is the source-drain current of the transistor gated by the voltage $V_{\text{gate}}$ and $\tau_{\text{LED}}$ is the response time of the LED. The transistor couples the two linear systems of receiver and transmitter and provides the system with a non-linearity, necessary for making use of the node device in neural networks. Furthermore, there is a non-linear contribution from the fact that the photoconversion efficiency of the LED is strongly dependent on the LED current as plotted in Fig. 3(c).

The output power of the node device is thus given as

$$P^{\text{out}}(t) = \frac{\hbar \omega}{e} I^{\text{LED}}(t) \eta(I^{\text{LED}}) \tag{3}$$

where $\hbar\omega/e$ is the fraction of the emitted photon energy over the elementary charge.

With the dynamical properties of single node laid down, the properties of a network of such devices can now be formulated. We describe our network as set of layers $L$ interconnected by weights $W$ and denote layer index by roman subscripts. For each layer $i$, the model voltages are calculated by integrating over time, using the input currents from all other connected layers as boundary conditions, according to

$$V(t) = \int_0^t dt' \, A_i V_i(t') + B_i I_i(t') \qquad (4)$$

where the input currents are given by the weighted output of the connected layers as

$$I_i^{in}(t) = \sum_j W_{ij} P_j^{out}(t) \qquad (5)$$

for each respective wavelength channel. For each node, Eq. (5) needs to be evaluated twice, one time for the inhibitory input current $I^{inh}$ and subsequently for $I^{exc}$, each computation with a different set of connecting matrices $W$ depending on the network architecture. In Eq. (5) the assumption of instantaneous transmission, as noted above, has been used.

In Eq. (4), the $A$ and $B$ matrices are layer specific, and the weight matrices $W_{ij}$ connect the node devices of layer $i$ to other layers $j$, taking account the optical coupling strength between the nodes. Once the voltages are known, the output power of the nodes of layer $i$ can be updated as

$$I_i^{LED}(t) = \int_0^t dt' \, \frac{1}{\tau_{LED}} \left[ I_i^{sd}\left(V_{gate}(t')\right) - I_i^{LED}(t') \right] \qquad (6)$$

under the assumption that the $I^{sd}$ responds instantaneously to changes in $V_{gate}$, as noted above. From these two equations the main system variables are calculated. If the network has recurrent connections, that is, if the input currents of layer $i$ depends on the output currents of the very same layer $i$ at an earlier time, the two equations need to be solved in parallel and cannot be separated as they could be in a feed-forward network.

As global boundary conditions to the network, an input layer with output powers $P^{out}$ known at all times connected by $W^{in}$ can be used to fully specify the network dynamics. Output layers can also be defined that records collected output currents with a set of specific outcoupling weight matrices $W^{out}$.

The resistances, capacitances, and transistor parameters needed to specify the model where obtained by dynamic finite element modelling of the semiconductor subcomponents, including the receiver NW photodiodes, the NW FET, and the NW LED. For the photodiode and FET, the dynamic finite element modelling was calibrated against experimental data in order to provide a realistic set of parameters (see SI for a table of all modelling parameters).

## III. APPLICATIONS

### A. Reservoir network

Physical reservoir networks do not require training of the internal network weights [35], often referred to as reservoir weights. Instead, the readout weights are trained to extract useful information from the reservoir states. To demonstrate how our device can be used as the fundamental building block for physical reservoir computing, we simulate a network of devices connected with random weights. One realization of such a system would be node devices placed at random on a substrate where their optical coupling strength, i.e. their connection weights, would be dependent on their relative orientation and distance.

As we are interested in physical reservoirs, we focus on a realistic situation where we fabricate devices that have different memory decay constants $\tau_{\text{mem}}$. We investigate how such a network performs compared to a network of identical components. In the following we will thus distinguish between *single-$\tau$* networks and *multi-$\tau$* networks. All random time constants, weights and frequencies are drawn from a uniform distribution.

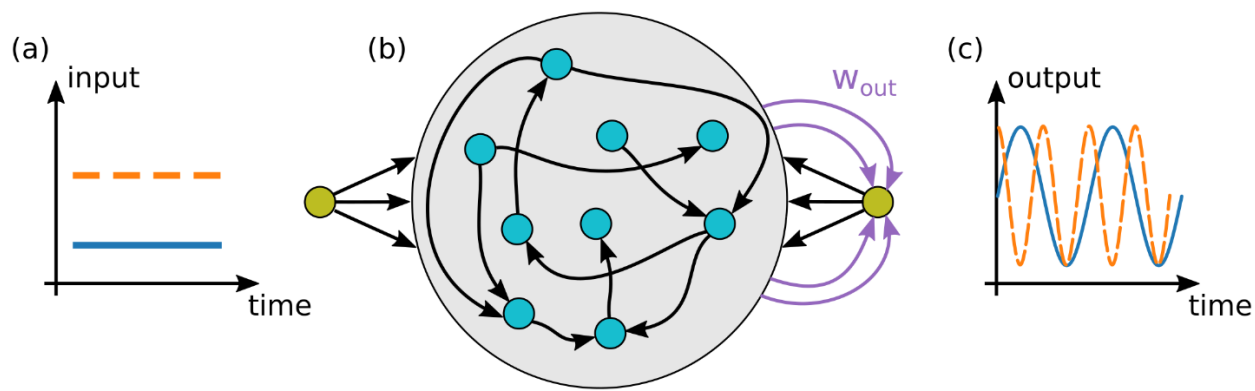

FIG. 5: Illustration of the reservoir network and the sine wave generation task it was trained to solve. (a) The network is fed a step-like signal in time through random input weights. (b) A reservoir of physical nodes with random weights generates rich internal dynamics because of random feedback connections from the output node. (c) Using teacher forcing the output weights are trained to provide a sine wave with frequency proportional to the input signal [36,37].

We train the network for a sine wave generation task, which is a classic test case for reservoir networks with a feedback connection [36,37]. As illustrated in Fig. 5, the network receives a piece-wise constant input (a) and should respond by generating a sine wave (c) with a modulated frequency proportional to the input signal amplitude. The input and target output signals are created by randomly dividing the time interval of the total simulation into segments $i$ and assigning to each a random frequency $f^i$ from the interval $[f_{\min}, f_{\max}]$. Then the signal $y_n$ at the discrete time $n$ can be expressed as

$$z_n = z_{n-1} + 2\pi f^i_{n-1} dt \tag{7}$$

$$y_n = \frac{1}{2}(sin(z^n) + 1) \tag{8}$$

where the time interval $dt$ is constant and chosen to properly resolve the sine wave at the maximum frequency $f_{\max}$.

The sine wave generation task requires a feedback loop from the output node back into the reservoir, as shown in Fig. 5(b), and the standard way of training a network for this task is by teacher forcing [38]. In this scheme, all weights for input, reservoir and feedback are randomly generated, while the readout weights $w_{out}$ are set to zero during training. The target signal is instead fed directly by the output node into the network via the feedback weights. After the training sequence, the optimal output weights $w_{out}$, are found using ridge regression [38].

The network was constructed with one input layer consisting of a signal node and a bias node, two hidden reservoir layers and a final output layer. Two hidden layers were used to allow for effective negative weights. Note that for the physical networks where signals are transferred optically by intensity, all weights are positive. In order to remedy this, the two hidden layers have different output wavelengths, but identical excitation and inhibition wavelengths. The inhibiting channel thus has an output wavelength that matches the inhibition wavelength of the excitatory layer, and vice versa. This way negative weights can be encoded in the network. Non-zero weights were generated randomly only for a limited number of outgoing connections from each node to ensure a specific sparsity of the weight matrix. These weights connect each layer recurrently with itself, as well as with the other hidden layer. No recurrent (diagonal) connections, i.e. from a node to itself, were allowed as constrained by the physical nature of the NW node device. The total weight matrix was scaled in order to set the spectral radius $\rho = 0.6$, as a value below 1.0 is needed for the network to fulfil the echo state property [38] (see Appendix D and SI for details). We restrict the output signal to only comprise of connections from the reservoir with excitatory output, that is, half the reservoir. This is done in order to have only a single wavelength in the output signal.

Training was carried out with reservoirs of 100 nodes and a sparsity of the network of 10%. Two different versions of the reservoir were used. On the one hand, an artificial network where all node devices share the identical memory decay times, as specified by the parameters in the dynamic matrix $A$ in Eq. (B1), and on the other hand, a more realistic network where the decay times vary due to variations in the device fabrication. In the second and more realistic version, the memory time decay $\tau_{mem}$ of the node devices were picked randomly from a uniform distribution on the interval $[0.1\tau_{mem}^0, 1.9\tau_{mem}^0]$, with $\tau_{mem}^0 = 5$ ns being the fixed memory time scale in this numerical experiment.

An example of a training sequence is shown in Fig. 6, where output signals of the single-tau and the multi-tau reservoirs are compared to each other in the upper panel (a), while the target output signal and the input control signal are plotted in the lower panel (b). Here, the random number generator used the same seed in order to ensure identical test environments. The first 2000 ns were used as a training sequence after which the output weights were calculated and set. For the rest of the sequence, the networks were then set free to continue the prediction without any supervision. It is clear from this example that the multi-tau network outperforms the single-tau network.

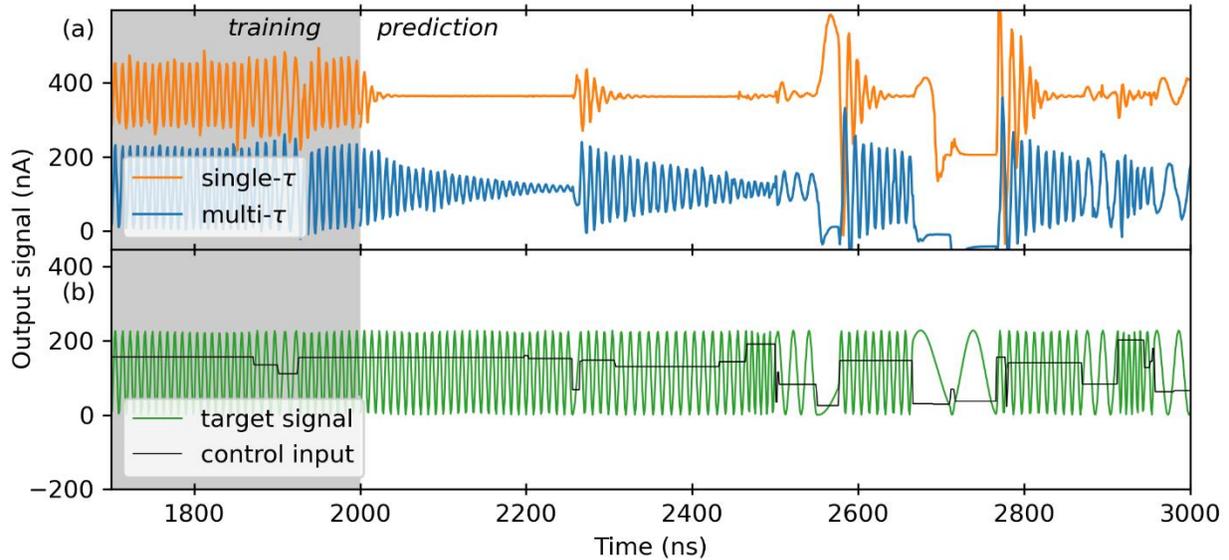

FIG. 6: Training and prediction sequence. Example of a network trained for 2000 ns after which the target output signal was replaced by the network output signal constructed by the trained weights $w_{out}$. (a) Comparison between two reservoirs with a single or distributed value of memory time constants $\tau_{mem}$. The *single*-$\tau$ line has been displaced for clarity. (b) Input control signal and target output signal for the training and prediction sequences.

To get quantitative results the error in the predicted frequencies were estimated. For each trial, the prediction sequence was divided into windows of 128 ns with a 32 ns overlap between the windows. The dominating frequency component inside each interval was calculated by a standard fast Fourier transform. Repeating the process for the target output signal, an estimate of the error could be found using the least squares of the errors of each sequence. In Fig. 7(a) we show the rms errors of 24 trials comparing the single-tau and multi-tau networks. Moreover, in Fig. 7(b) we show the errors without taking the mean for each simulation, instead summing the individual errors for all time windows used in the spectral analysis. Using the $5\Delta f$ error limit as a threshold, we can conclude that 75% of the multi-tau networks are successful, while only 20% of the single-tau networks perform on the same level (although close to the error threshold). From the results there is a distinct advantage in having a distribution of memory constants for the sine wave generation task.

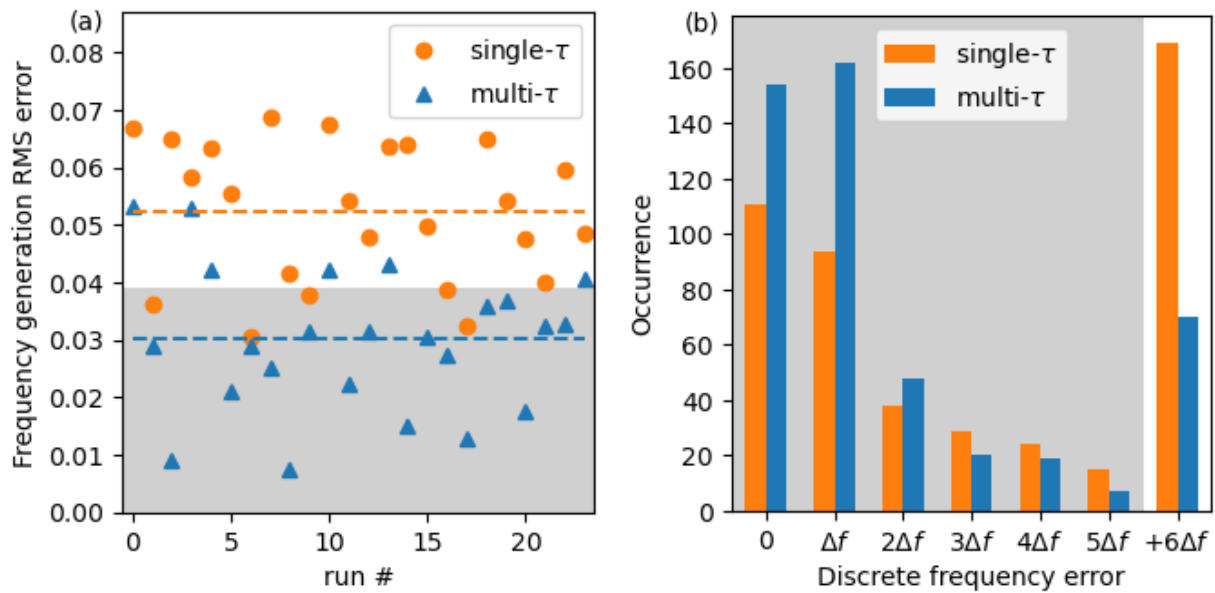

FIG. 7: Prediction error statistics. (a) RMS errors in predicted frequencies for 24 runs with randomly generated networks and target signals. Mean error is shown as a dashed line and the shadowed area corresponds to an error less or equal than $5\Delta f$ where $\Delta f = 0.08$ GHz is set by the discrete frequencies from the Fourier spectral analysis of the time signal. (b) Histogram over the accuracy of the predictions in all time-windows used in the spectral analysis of the time signals of the 24 runs of (a). The shaded region corresponds to the same tolerated error as in (a).

### B. Biologically constrained neural network

As a second network system, we choose a neural circuit model based on the central complex in the insect brain, able to perform homing through path integration – a special form of navigation typical to insects [8] which lends itself well to implementation with NW components [28]. In this section we show that a complete navigational network can be built using our short-term memory components and study the requirements on accuracy and repeatability.

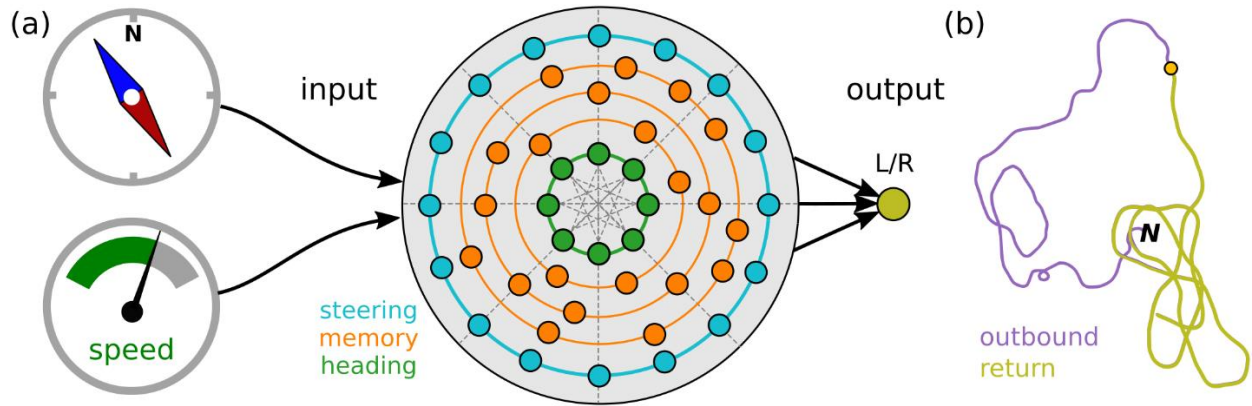

FIG. 8: Navigational network. (a) Schematic diagram of our implementation of the recurrent network of [8]. Inputs to the network are optical flow as a measure of speed and the perceived heading with respect to an internal compass. The nodes of the innermost layer for heading (TB1 cells) are interconnected recurrently with weights in a sinusoidal pattern. Orange indicates the memory layers (rectifier, CPU4 and pontine cells) and the blue the steering layer (CPU1 cells). (b) Example result for an outbound journey from the nest (N) to the yellow dot and back to N of a duration of 1500 ns.

A graphical description of the modelled network is presented in Fig. 8. The main task of the network is to calculate the steering signal during homing, in order to lead an insect – the agent – towards its nest after a foraging trip. The steering signal is represented by the activity in the two groups of CPU1 cells and their summed activity will determine whether the agent turns right or left and how strongly. The groups of CPU1 cells are informed by the memory CPU4, the memory balancing activity of the pontine cells and the heading direction stored by the TB1 cells. The TB1 layer holds the heading direction of the agent and provides output to both the memory (CPU4) cells and the steering (CPU1) cells. The pontine cells provide a rebalancing of the memory signals of the left and right hemisphere. These cells are important if the flight is holonomic to some degree, meaning that the head direction is different than the ground velocity, due to factors like wind or aggressive steering. Finally, the memory cells in the CPU4 layer integrate the travelled route in order to store the direction home as a population code [8], by summing the activity from the optical flow input and the head direction cells TB1. To do this, it is important that the nodes representing the CPU4 layer have a memory timescale long enough to record the outbound travel. By tuning the built-in short-term memory time constant of our device, it can meet these requirements and we can thus construct all network layers using our device design. In addition, it is possible to tune the slope and shift of the activation function by adjusting the transistor properties (see SI for details).

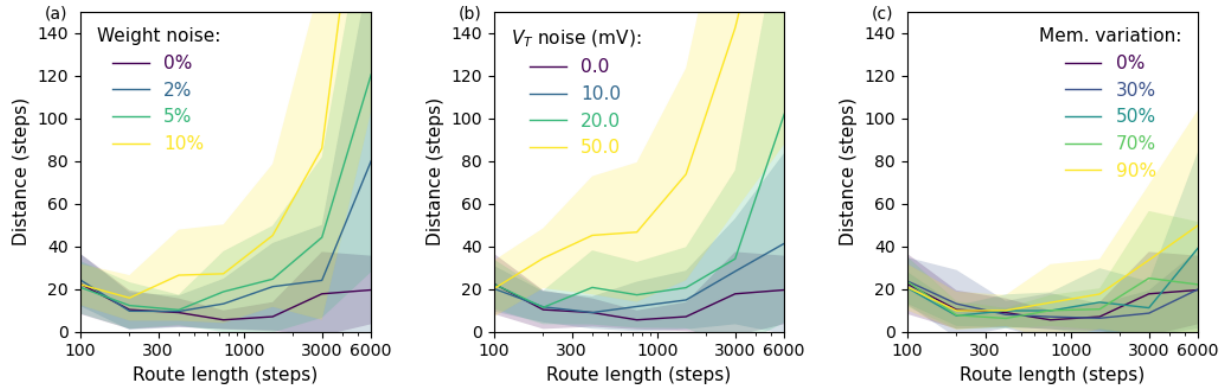

FIG. 9: Statistical studies of the robustness of the navigational network with respect to distortions and errors. Minimum distances to the nest are shown as function of outbound travel length. Parameters varied are weights connecting the layers (a), the threshold voltage of the transistors (b) and the variation in the memory time constant of the CPU4 layer (c).

To test whether our NW-based device of Fig. 2 could perform as the fundamental unit of the navigational network, we simulated the system using physical devices for the nodes of the TB1, CPU4, Rectifier, Pontine and CPU1 layers, essentially all nodes except the input layers TN2, TL and CL1. The rectifier layer is introduced here to perform a non-negative clipping of input activity to the CPU4 layer, which was done artificially in [8]. The exact same connection pattern as in [8] was supplied and a few parameters of the biological model were adjusted to our continuous time simulation. The nodes representing the CPU4 memory cells were tuned to have a long memory constant, $\tau_{mem} = 100$ µs, compared to the rest of the network where $\tau_{mem} = 1$ ns. For the activation functions of the layers, where Stone et al. used specific slope and bias parameters for the individual layers, we use only two different distinct set of parameters, which can be thought of as two alternative ways of transistor processing. For the CPU1 layers, a bias is introduced relative to the other layers, shifting the activation function by 0.2 V. Such a change could be realized in many ways, for example by changing the metal on the transistor gate electrode (see SI). Apart from this shift, all layers share the same activation function. The input signals from layers TN2, TL an CL1 were generated as in [8] with a typical added noise of 20%. To compare to the simulations of [8] where an arbitrary time step was used, we use 1 ns as a comparative time step. This means that an outbound flight of 1000 steps have a duration in our simulation of 1 us.

The model was tested over a large number of trials to statistically check the performance. An example route can be seen in Fig. 8(b) where the agent switches from a random outbound journey to a homing behaviour at the point of return. This is followed first by a more or less straight line towards the nest and then the agent reverts to a search behaviour when it is close enough to the nest. This is a feature of the model and the effect comes from the continuous update of the memory, meaning that the direction towards the nest is updated during the entirety of the flight.

In Fig. 9 we show the results of three different set of trails, where different types of noise and distortions were introduced in the mode. Before going into the details of each set of trials, we note that the undistorted network without any noise (dark purple line) performs well up to outbound travels up to 6000 steps, with an average distance to the nest (upon return) of only 20 steps. Panel (a) describes the impact of adding noise in the weights between the layers. Keeping the device perspective, these weights can be affected by for example, absorption efficiency in the receiver NW, external quantum efficiency in the transmitter LED and deviation in distances between two communicating nodes. The results indicate that such errors can be up to a few percent in order for the network to be able to complete its task accurately, a result similar to what was found in [8]. Deviations were also introduced in the transistor threshold voltages and the effect is shown in panel (b). These errors effectively made transistors in each layer different by assigning a random threshold voltage around the fixed value, from a normal distribution with standard deviations as shown in the legend of panel (b). From this study we can conclude that deviations need to be limited to around 20 mV to preserve good navigation capabilities for 1500 step journeys. Finally, we study the deviations in the memory time constants for the CPU4 memory layer, with the results in panel (c). Here we note, in contrast to the two previous trials, that the network is not so sensitive to the exact value of this time constant. This basically means that the time constant needs to be longer than those of the rest of the network, but the exact value is unimportant.

## IV. Conclusion

In the present work we have explored the concept of an artificial neuron with an internal (leaky) memory. Inspired by biological models this can be important to achieve functionality in e.g. insect brain derived navigation circuits as well as more artificial networks such as the reservoir networks. We present a specific optoelectronic hardware solution using III-V semiconductor NWs and prove the functionality using experimentally benchmarked parameters in the simulations. The proposed NW component is a further development of an initial circuit that included light interconnectivity and analog sigmoid signal processing. The addition of a memory can be seen as a modular extension of this artificial neuron that does not add additional complexity in terms of network connections. The fundamental concept and the added functionality demonstrated here should be widely applicable for a variety of hardware solutions. Interestingly, we find that the variability in the time constant of the memory retention, which would naturally occur in most hardware solutions, can be used to significantly enhance the functionality of certain types of networks. Network heterogeneity and its possible benefits have been studied also for spiking neuron models [39]. While the learning process of neural networks has mostly been explored for the interconnects, for example optimization by the backpropagation algorithm, it will be interesting to further explore the in-neuron memory which is relatively easy to implement.

The system implemented here does not have an external clock, as the timing is driven by the input signal speed. In the present case we have focused on the high-speed limit of our system which is set by the LED on/off time that is in the ~1 GHz range. While the natural timescales are much slower, the present network can work efficiently at much higher speeds. For calculation purposes these high speeds are relevant for extreme performance, just as the human brain project performs biological simulations at an artificially high speed [40], however for responding to a natural situation such as mini drone navigation one would potentially slow the system down. To accomplish this the memory retention time needs to be significantly increased. This could be done both using nonvolatile NW based designs [41–43] or more traditional Si charge trapping devices that can be connected to the NWs [44,45]. As the memory is only included in the artificial neurons, this construction potentially using a standard chip platform is possible.

The statistical tests in Fig. 9 show estimates on how precise the fabrication of the physical devices needs to be. Similar to [8] we find that weight noise cannot exceed 5%, see Fig 9(a), for a reasonable accuracy of the network. In addition, the transistor threshold values are equally important, as shown in Fig 9(b), where deviation above 10 mV are problematic to the network. On the other hand, the precise value of the memory time constants in the different memory nodes of layer CPU4 is not critical, as shown in Fig 9(c). These estimates will be valuable for future device fabrication, providing guidance on the impact of different errors on the node device performance.

For the implementation of the insect-inspired network of [8], the memory timescale in the CPU4 layer needs to be significantly longer than those in the other layers. In addition, each activity update needs to be small in order not to exceed the memory capacity, effectively requiring a small weight. In our physical device, the increased memory timescale and decreased weight value can be simultaneously achieved by increasing the storage capacitance $C_{\text{MEM}}$. In addition to memory longevity, this produces a smaller voltage change per added charge, resulting in smaller weights.

Comparison to other neuromorphic and CMOS systems has previously been deemed favorable for light interconnects [28]. The main energy loss was previously identified to be in the amplification of the signals and the LED which are still the same.

## Acknowledgments

The authors thank Barbara Webb for many useful discussions and for allowing us to reuse part of the code from [8] for the biological network simulations. This work was supported by the Swedish Research Council,

NanoLund, the Office of Naval Research (Grant No. N62909-20-1-2038) and the European Union Horizon Europe project InsectNeuroNano (Grant 101046790).

## Appendix A: Calculation of the stationary activation function

Here we detail the calculations of the example activation function plotted in Fig. 4(b). The figure shows the stationary solution of VGATE as a function of constant excitation current, keeping the inhibition current at zero. To solve the stationary version of the receiver circuit (left column) in Fig. 2(b), we neglect the saturation current of the reverse-biased inhibiting photodiode and express the current through the memory resistance as

$$I_{\text{Rmem}} = I_p - I_0(\exp(V_{\text{exc}}/nk_BT) - 1) \tag{A1}$$

where $I_p$ is the photoinduced current and

$$V_{\text{exc}} = I_{\text{Rmem}}(R_{\text{exc}} + R_{\text{mem}}) \tag{A2}$$

Inserting, we obtain the equation

$$I_{\text{Rmem}} = I_p + I_0 - I_0 \exp(I_{\text{RMEM}}(R_{\text{exc}} + R_{\text{mem}})/nk_BT) \tag{A3}$$

which can be solved for $I_{\text{RMEM}}$ using the Lambert $W$ function resulting in

$$I_{\text{Rmem}} = I_p + I_0 - \frac{nk_BT}{R_{\text{exc}} + R_{\text{mem}}} W\left(\frac{I_0(R_{\text{exc}} + R_{\text{mem}})}{nk_BT} \exp\left(\frac{(I_p + I_0)(R_{\text{exc}} + R_{\text{mem}})}{nk_BT}\right)\right) \tag{A4}$$

providing the current through the memory resistance as a function of photocurrent. This is then directly related to the gate voltage $V_{\text{gate}} = I_{\text{Rmem}} R_{\text{mem}}$. As seen in Fig. 4(b), the gate voltage saturates when the photodiodes threshold is reached. Plugging the gate voltage into our typical transistor function yields the full activation function relating input to output current. In summary, the photodiode and transistor threshold combine to provide an almost sigmoidal shape of the stationary activation function.

## Appendix B: Dynamical model parameters

Here we provide the full matrices for the dynamic model of Eq. (1). The matrix $A$ describes the voltage dynamics in the receiver part of the circuit,

$$A = \begin{pmatrix} -1/R_{s,inh}C_{inh} & 0 & 1/R_{s,inh}C_{inh} \\ 0 & -1/R_{s,exc}C_{exc} & 1/R_{s,exc}C_{exc} \\ 1/R_{s,inh}C_{mem} & 1/R_{s,exc}C_{mem} & -1/C_{mem}(1/R_{inh} + 1/R_{exc} + 1/R_{mem}) \end{pmatrix} \quad (B1)$$

consisting of the different $RC$ constants of the receiver and memory part of the circuit diagram in Fig. 2(b). In the second term of Eq. (1), $B$ is a scaling matrix taking the capacitances into account as

$$B = \begin{pmatrix} -\dfrac{1}{C_{inh}} & 0 & 0 \\ 0 & \dfrac{1}{C_{exc}} & 0 \\ 0 & 0 & 0 \end{pmatrix} \quad (B2)$$

where the last diagonal element is naturally zero as there is no current source linked to the gate voltage $V_{gate}$.

Modelling the receiver as a linear system effectively means that the photodiode components, $D_{inh}, D_{exc}$, are excluded from the model. This is justified as the model in general is operated at voltages far from the diode threshold currents. The diode behaviour enters only as negative and positive limits on the $V_{gate}$, respectively,

$$-V_{inh}^{max} < V_{gate} < V_{exc}^{max} \quad (B3)$$

We model the efficiency of our nanowire LED by the ABC model under the approximate assumption that the carrier density in the active region of the LED is proportional to the current. This gives a direct relation between efficiency and current as

$$\eta(I) = \frac{A}{A + BI + CI^2} \quad (B4)$$

Where the parameters $A$, $B$ and $C$ model trap-assisted, radiative and Auger recombination, respectively. The effective parameters $B/A$ and $C/A$ were fitted to electrical modelling of our nanodevice design, as exemplified in Fig. 3(c).

**TABLE B1. Model parameters used in the dynamical simulations.** In order to tune the memory time constant, $R_{mem}$ and $C_{store}$, marked by asterisks, can be adjusted. The listed values correspond to a memory time constant of 1 ns.

| Parameter | Value |
|---|---|
| $R_{inh} = R_{exc}$ | 400 kOhm |
| $C_{inh} = C_{exc}$ | 50 aF |
| $\tau_{LED}$ | 90 ps |
| $R_{mem}$* | 1 MOhm |
| $C_{store}$* | 900 aF |
| $C_{gate}$ | 10 aF |
| A/B | 2.2 uA |
| C/B | 3.3 1/uA |

## Appendix C: Scaling of the signals in the network

The networks designed and tested in this work are optically connected via a broadcasting strategy. As there are no designated channels for signal transfer, most of the signal is lost, by design, and only a fraction is detected on the receiving side. Hence, most weights are typically a fraction of a percent and the gain required to maintain the signal is supplied inside each node by the current source driving the transistor and LED. To maintain a common weight structure, the internal gain and the transmission loss is balanced in our implementation.

As a first step we calculate a *unity coupling* coefficient $\theta$ which estimates the coupling strength needed between two node devices in order to transmit a signal without losses. It is basically the inverse of the low frequency electronic gain of the node device and we estimate it by the following expression

$$\theta = \frac{I_{exc}^{max}}{\eta(I_{LED}^{max})I_{LED}^{max}} \tag{C1}$$

where

$$V_{gate}^{max} = V_{exc}^{thres} \frac{R_{mem}}{R_{mem} + R_{exc}} \tag{C2}$$

$$I_{exc}^{max} = V_{exc}^{thres} \frac{1}{R_{mem} + R_{exc}} \tag{C3}$$

$$I_{LED}^{max} = IV_{FET}(V_{gate}^{max}) \tag{C4}$$

These expressions have been derived using the steady state version of Eq. (1) under the assumption of zero inhibition current. All weights in the system are thus scaled by $\theta$, allowing us to work with values around unity to describe out networks, while input signals, like the frequency modulated wave in Fig. 6(b), is scaled by $I_{\text{exc}}^{\max}$ when it enters the actual physical system. In most cases, the network devices are not all the same in terms of transistor characteristics and resistances, but as a scaling factor Eq. (C1) is perfectly adequate.

## Appendix D: Creation of the reservoir network

To create a balanced reservoir with weights that are effectively both positive and negative, even though output is measured in intensity, the reservoir is split into two partitions $R+$ and $R-$, where the sign indicates excitatory or inhibitory output wavelength. The resulting weight matrix connecting the hidden layer has the following structure:

$$W_{\text{res}} = \begin{pmatrix} W_{++} & W_{-+} \\ W_{+-} & W_{--} \end{pmatrix} \quad (M14)$$

Acting upon the set of reservoir nodes $R = (R_+, R_-)$ at each update. In the matrix $W_{\text{res}}$ diagonal elements are zero as the nodes are not allowed connect to themselves. The spectral radius of the network is calculated using the matrix

$$W'_{\text{res}} = \begin{pmatrix} W_{++} & -W_{-+} \\ W_{\pm} & -W_{-} \end{pmatrix} \quad (M15)$$

Which provides the actual impact of the weights on the node devices. The spectral radius is evaluated as the eigenvalue of maximum amplitude $r = \max|\lambda|$ of this matrix and the elements of $W_{\text{res}}$ are then scaled by $r_{\text{spectral}}/r$.